%% file: paper.tex
\documentclass[runningheads]{llncs}

\begin{document}
\title{A Survey on Sustainable Software Ecosystems to Support Experimental and Observational Science at Oak Ridge National Laboratory%
\thanks{This manuscript has been authored by UT-Battelle, LLC, under contract DE-AC05-00OR22725 with the US Department of Energy (DOE). The publisher acknowledges the US government license to provide public access under the DOE Public Access Plan (\url{https://energy.gov/downloads/doe-public-access-plan}).}
}
\titlerunning{Survey Software Eco EOS ORNL}
%

\author{David E Bernholdt\orcidID{0000-0001-7874-3064} \and \\
        Mathieu Doucet\orcidID{0000-0002-5560-6478} \and \\
        William F Godoy\orcidID{0000-0002-2590-5178} \and \\ 
        Addi Malviya-Thakur\orcidID{0000-0002-2681-9992} \and \\
        Gregory R Watson\orcidID{0000-0002-8591-2441}
        \thanks{These authors contributed equally to this work.}
}

\authorrunning{D.E.~Bernholdt, M. Doucet, et al.}
%
\institute{
Oak Ridge National Laboratory \\ 
Oak Ridge TN, 37830 USA \\
\url{https://www.ornl.gov}
}
\maketitle              
\begin{abstract}
In the search for a sustainable approach for software ecosystems that supports experimental and observational science (EOS) across Oak Ridge National Laboratory (ORNL), we conducted a survey to understand the current and future landscape of EOS software and data. This paper describes the survey design we used to identify significant areas of interest, gaps, and potential opportunities, followed by a discussion on the obtained responses. The survey formulates questions about project demographics, technical approach, and skills required for the present and the next five years. The study was conducted among 38 ORNL participants between June and July of 2021 and followed the required guidelines for human subjects training. We plan to use the collected information to help guide a vision for sustainable, community-based, and reusable scientific software ecosystems that need to adapt effectively to: i) the evolving landscape of heterogeneous hardware in the next generation of instruments and computing ({\it e.g.} edge, distributed, accelerators), and ii) data management requirements for data-driven science using artificial intelligence.

\end{abstract}
\section{Introduction}
Computational science and engineering (CSE) is well-established as a compute, data, and software-intensive approach to scientific research with decades of history behind it. Traditionally, the software and data aspects of CSE have been marginalized, as incentive systems have emphasized novel scientific results over considerations related to the software and data that underpin them or indeed the skills and expertise of the people who develop the software.  However, in recent years, there has been a trend to improve this through an increased understanding of the importance of the software and data to achieve high quality, trustworthy, and reasonably reproducible scientific results.

Experiment and observation have a far longer history in the conduct of scientific research than computationally-based approaches. As computing has become more capable and accessible, experimental and observational science (EOS) has also progressively expanded its use of software and computing for instrument control, data collection, reduction, analysis; data management; and other activities. It would not be a stretch to say that modern EOS relies on software as much as computational science, but with differences in complexity and scale. Therefore, as new approaches in instrumentation, and enhanced computational capabilities make feasible novel and more complex experiments, the reliance on software and computing in many areas of EOS is growing rapidly~ \cite{Barreiro_Megino_2017,LIGOSpecialIssue,RICHABBOTT2021100658,b1e0c69119bf4a7cb524b242b511a282}. In addition, EOS researchers are increasingly seeking to harness high-performance computing (HPC) that had historically been the province of computational scientists to deal with rapidly increasing data volumes. Modeling and simulation techniques originated in the CSE community are widely used to help understand and interpret EOS data~\cite{Barreiro_Megino_2017,9307775}. As a result, the EOS community is transforming towards an increasing software- and compute-intensity level.

This paper represents the results of a small survey targeting EOS-focused staff within one organization that is focused heavily on EOS: Oak Ridge National Laboratory (ORNL). ORNL is the largest multi-program laboratory  under the U.S.~Department of Energy (DOE) Office of Science. The survey attempts to identify challenges facing EOS software stakeholders and how they anticipate the situation evolving over a five-year time frame. The structure of the paper is as follows: Section~\ref{sec:background} presents background information on the need for software ecosystems in related fields, while a brief description of the current landscape of software ecosystems in scientific computing is also presented in anticipation of the survey responses. Next, an overview of the survey structure, methodology and questions is described in Section~\ref{sec:survey}. Next, results from the survey and a discussion is presented in Section~\ref{sec:survey_results}. This section is perhaps the most important in the report as we attempt to craft a narrative and interpretation by correlating the overall answers. Lastly, our conclusions from the survey results and analysis are presented in Section~\ref{sec:conclusions} outlining the most important takeaways from the collected data. 

\input{relatedwork}

\input{survey}

\input{survey_results}

\section{Conclusions}
\label{sec:conclusions}
Our survey attempts to provide empirical evidence to understand the landscape of the software supporting EOS activities at a major research laboratory. Results suggest that the field is still focused on classical computing approaches. More recent developments in computing that are the norm outside of science, like AI, edge computing, and cloud-based services, are still not being used extensively in existing projects. On the other hand, the field is clearly moving towards a broad collaborative approach. Projects tend to be multi-institutional and benefit a wider array of users. At the same time, most projects are comprised of a handful of developers, and most project interactions tend to be between a few developers. This highlights the niche nature of highly-specialized scientific software. We have also seen that data dissemination and sharing are becoming a focus across a range of scientific endeavors. In addition, projects feel confident that they will tackle challenges in the next five-year time frame. We understand these results as a reflection of the alignment to sponsors' expectations, for which its impact on ORNL EOS deliverables must justify investments in aspects of scientific software shown in this survey. In the future, we would like to use the insights gained from this survey to better support EOS developers and develop guidelines for a sustainable EOS software and data ecosystem. 

\bibliographystyle{splncs04}
\bibliography{paper}

\end{document}

%% file: relatedwork.tex
\section{Background}
\label{sec:background}

Software ecosystems are an increasingly important component of scientific endeavors, particularly as computational resources have become more available to include simulations and data analysis in research~\cite{manikas2013software}. 
Nevertheless, there is still debate of what constitutes a ``software ecosystem".
Dhungana {\it et al.}~\cite{dhungana2010software} points out the similarities between software and natural ecosystems. Monteith {\it et al.}~\cite{monteith2014scientific} indicates that scientific software ecosystems incorporate a large environment that includes not only software developers, but also scientists who both use and extend the software for their research endeavors. Therefore, it is important to acknowledge the different needs and goals of these groups when considering the broad breadth of scientific software. Hannay {\it et al.}~\cite{5069155} asserts that there is a great deal of variation in the level of understanding of software engineering concepts when scientists develop and use software.Kehrer and Penzenstadler ~\cite{DBLP:conf/re/KehrerP18} explore individual, social, economic, technical and environmental dimensions to provide a framework for sustainability in research software.
As described by Dongarra {\it et al.}~\cite{doi:10.1177/1094342010391989}, the high-performance computing (HPC) community identified in the last decade the need for an integrated ecosystem in the exascale era. Efforts resulted in the DOE Exascale Computing Project (ECP)~\cite{ecp:www}. Within the scope of ECP, the Extreme-scale Scientific Software Stack (E4S)~\cite{heroux2019extreme} aims to reduce barriers associated with software quality and accessibility in HPC.

Few studies have been carried out to assess EOS software ecosystems, so it is necessary to draw from experiences in other communities. A critical part of a software ecosystem are the developers themselves, since they play a crucial role that requires establishing sustainable collaborative relationships within the community. Sadi {\it et al.}~\cite{10.1007/978-3-319-19243-7_17} draws from test cases on mobile platforms to provide an in-depth analysis of developers' objectives and decision criteria for the design of sustainable collaborations in software ecosystems. The guidelines to pursue requirements for a software ecosystem is discussed by Kaiya~\cite{KAIYA20181243} to allow for a decrease in effort, increase
of gain, sustainability and increase in participation of developers in an engaging conversation.  Empirical assessments of how modern software and data practices are used in science are provided in recent studies ~\cite{HEATON2015207,lamprecht2020towards,storer2017bridging}. Software ecosystems can be vast and have been continuously evolving through the development of new processes, inventions, and governance~\cite{Bartlett2017}. Therefore it is important that scientific communities develop a tailored plan that relies on software reuse and development of an interdependent~\cite{6676899} and component-centric~\cite{10.1145/1842752.1842782} software ecosystem that understands the goals and needs within the context of EOS.  

%% file: survey.tex
\section{Survey Overview}\label{sec:survey}

\subsection{Survey Motivation}
As a preeminent scientific research organization, ORNL uses and creates a great deal of the software used to undertake CSE. Software is central to modern EOS for data acquisition, reduction, analysis, distribution, and related modeling and simulation used for CSE. However, instruments and sensors are growing more capable and sophisticated and experiments more complex.  At the same time, the computing landscape is also changing, with a transition towards ``edge" systems that are situated closer to the experiments, and the increasing use of cloud and HPC resources.  These factors drive complexity in software and consequently create greater challenges for developers. In order to better understand this changing landscape, we decided to conduct a \textit{Software Ecosystem Survey}. The goal was to collect a range of information about immediate and future development needs, the environmental factors influencing these needs, and the skills and training necessary for developer teams to be able to effectively work in these changing environments. We hope to use the information collected from the survey to improve participation and collaboration in software ecosystems and aid in creating sustainable software as we enter an exciting new frontier of scientific discovery.

\subsection{Survey Design}

The survey is organized into 10 major sections for a total of 34 questions. Responses are completely anonymous and meet the requirements for conducting ethical human-subject research studies. The survey design was focused on multiple-choice questions and answers, allowing the taker to register quickly and efficiently. The survey is exhaustive and might take up to 45 minutes to complete. 

The survey begins with the description and motivation so that participants understand what the survey is trying to achieve as well as what information is expected from them. The survey then asks a series of questions to provide background information and project demographics to understand the nature of the software project and its contributors. A series of questions relating to the technical approach used by the project are asked in order to understand current software and data needs and challenges. Next, there are questions to understand the current skill levels and how new skills are acquired among software project participants. Finally, questions about the future demographics, future technical approach, and preparations for the future are formulated to understand the leading technological disruptions and the significant challenges projects will be facing in the next five years. In addition, we formulated questions on confidence to address current and future challenges at a personal and team level. The survey was developed using Google Forms\textsuperscript{TM} and the responses were recorded over the course of a few weeks. 

\subsection{Selection of Participants}
We invited individuals at ORNL working across diverse scientific domains developing research software to participate in this survey. This included developers from nuclear energy, biostatistics, transportation, building technology, geospatial analytics, among several others. Almost all of the invited participants had experience developing scientific software and were part of the broader software development community. The participants' anonymity was maintained by excluding any personal or work related information.

%% file: survey_results.tex
\section{Survey Results}
\label{sec:survey_results}

We received a total of 38 responses to the survey (raw results available at \cite{raw_data}). None of the questions were required, so the number of responses to specific questions may in some cases, be fewer, which we will indicate as necessary.  For questions about the project demographics and technical approach, we asked both about the current situation and the situation expected five years from now.

\subsection{Background Information}
This survey section consists of three questions to identify the respondents' relationship with scientific software activities and roles. The first question attempts to classify the respondents' work in relation to software. Overall, 71\% of respondents indicated that research software is a primary product of their work, while 29\% indicated it is not.
The other two questions asked respondents to characterize the portion of their work time spent {\it using} or {\it developing} research software. Table~\ref{tab:time} shows that 67\% of the respondents spend at least 41\% of their time as developers. We take this as a good indication that the survey respondents are in our target audience. 

\begin{table}[tb]
\centering
 \caption{Responses to time percentage spent {\it using} or {\it developing} software.}
\label{tab:time}
\setlength{\tabcolsep}{0.5em}
\begin{tabular}{lccccc} 
 \hline\hline
 Research Software             &   & Time percentage &  &  \\
 Activity   &  0-20\% & 21-40\% & 41-60\% & 61-80\% & 81-100\% \\
\hline 
 Using      & 34.2\% & 23.7\% & 18.4\% & 13.2\% & 10.5\% \\
 Developing & 21.2\% & 13.2\% & 23.7\% & 13.2\% & 28.9\% \\
\hline\hline
\end{tabular}
\end{table}

\subsection{Project Demographics}
In the remainder of the survey, we asked respondents to focus on the single research software development project that they considered most significant in their work, as the Project Demographics section of the survey was intended to characterize aspects of their particular project. Overall, 82\% of respondents reported that they were users of the software as well as developers, whereas 18\% were exclusively developers.  This is consistent with our informal observations of computational science and engineering, where the majority of developers are also users.  Looking five years out, respondents had essentially the same expectations (81\% and 19\%).

Table~\ref{tab:size} illustrates responses to questions about the size of the development team, in terms of the overall number of active developers on the project and the number of those developers the respondent interacts with on a regular (weekly) basis.  The results show that roughly two-thirds (58\%) of projects are comprised of no more than 3 developers, and even on larger projects, the majority (68\%) regularly interact with no more than 3 team members.  However, there are also a considerable minority of software project teams (18\%) with more than 10 developers.

\begin{table}[tb]
\centering
 \caption{Number of developers in a project and how many of these interact weekly with the respondent.}
\label{tab:size}
\setlength{\tabcolsep}{0.5em}
\begin{tabular}{p{0.25\columnwidth}lcccccc} 
 \hline\hline
              &   & People &  &  \\
Developers that are:    &  0 & 1 & 2-3 & 4-5 & 6-10 & $>$ 10 \\
\hline 
 Active including \\
 the respondent     & n/a & 10.5\% & 47.4\% & 18.4\% & 5.3\% & 18.4\% \\
 \vspace{0.1em}
 Interacting weekly \\
 with the respondent & 7.9\% & 21.1\% & 39.5\% & 18.4\% & 10.5\% & 2.6\% \\
\hline\hline
\end{tabular}
\end{table}

We also tried to characterize the organizational breadth of the project teams. Table~\ref{tab:breadth} shows that the majority of projects are of multi-institutional nature. Of the project teams comprised exclusively of ORNL staff members, the largest number included staff from multiple directorates (the largest organizational level at ORNL), but the next largest number included staff from a single group (the smallest organizational level). Our informal observation is that in CSE at ORNL, the majority of projects are also multi-institutional. This is an indicator that while some of the software serves ORNL specific scientific purposes, a large portion exposes the team to outside organizations that could help leverage common software development activities via collaboration.  It is interesting to observe that in five years, there are expectations for a strong shift towards more broadly based project groups, particularly multi-institutional teams.

\begin{table}[tb]
\centering
 \caption{Organizational breadth of research software projects.  Organizational units within ORNL are listed from smallest (group) to largest (directorate). The ``ORNL" response denotes teams spanning multiple directorates.}
\label{tab:breadth}
\setlength{\tabcolsep}{1em}
\begin{tabular}{ llcc } 
 \hline\hline
 Breadth & Typical & Current & Expected \\
 all developers within  & Size & Percentage  & in Five Years   \\
 \hline
 Group & 8-10 staff & 13\% & 8\% \\ 
 Section & 3-4 groups & 8\% & 0\% \\ 
 Division & 3-4 sections & 0\% & 3\% \\
 Directorate & 2-4 divisions & 8\% & 3\% \\ 
 Whole lab &  8 science directorates & 16\% & 3\% \\
 Multiple institutions  & & 55\% & 84\% \\
 \hline\hline
\end{tabular}
\end{table}

The DOE Office of Science defines a user facility as ``a federally sponsored research facility available for external use to advance scientific or technical knowledge."\footnote{\url{https://science.osti.gov/User-Facilities/Policies-and-Processes/Definition}}  The DOE stewards a significant number of the user facilities in the United States, nine of which are hosted at ORNL\footnote{\url{https://www.ornl.gov/content/user-facilities}} and were targeted in our distribution of the survey.  We asked respondents whether their research software was intended for use at a user facility (whether at ORNL, within the DOE system, or elsewhere).
Only 8\% of respondents indicated that their software did \emph{not} target a user facility. The remainder indicated that the software was used in their own work at a user facility (13\%), or by multiple users (47\%).  A quarter of respondents (26\%) indicated that their software was part of the software suite that the facility offers to its users.

Finally, we asked the respondents to rate the importance of various software characteristics as summarized in Table~\ref{tab:9_25_characteristics}. The importance of most of these characteristics is expected to be about the same looking out five years, except that portability drops and security increases in importance. Given an opportunity to suggest additional ``moderate importance or higher" characteristics, we received 7 responses (some multiple listing characteristics).   We consider that some of the responses could be consolidated into the original list of characteristics, while others would be new additions.  The free-form responses are characterized in Table~\ref{tab:10_characteristics}.

\begin{table}[tb]
\centering
\caption{Characteristics considered to be of moderate or higher importance to the software projects.}
\label{tab:9_25_characteristics}
\setlength{\tabcolsep}{1em}
\begin{tabular}{lcc}
\hline\hline
 & Current & Expected \\
Characteristic & Percentage  & in Five Years \\
\hline
Functionality & 97\% & 97\% \\
Usability	& 87\% & 89\% \\
Maintainability or Sustainability	& 87\% & 89\% \\
Performance	& 89\% & 92\% \\
Portability	& 53\% & 34\% \\
Security & 21\% & 34\% \\
\hline\hline
\end{tabular}
\end{table}

\begin{table}[tb]
\centering
\caption{Analysis of free-form responses asking for additional important characteristics not included in the original six.  The authors consider most of the proposed characteristics could be included in one of the original six, while two could not.}
\label{tab:10_characteristics}
\begin{tabular}{lcl}
\hline\hline
Response & Occurrences & Original Characteristic \\
\hline
Unique capabilities & 1 & Functionality \\
Documentation & 2 & Usability \\
Intuitive & 1 & Usability \\
Robustness & 1 & Usability \\
Extensibility & 1 & Maintainability or Sustainability \\
Accuracy & 1 & \emph{n/a} \\
Correctness & 1 & \emph{n/a} \\
\hline\hline
\end{tabular}
\end{table}

\subsection{Project Technical Approach}

This section included six questions related to the technical aspects of the project and its environment.

First, we asked respondents to indicate the various technical categories applied to their software project.  Multiple categories could be selected, and a free-response option was allowed so respondents could add categories not included in the original list. Table~\ref{tab:11_27_categories} summarizes the results.  Not surprisingly, for a survey focused on EOS, data processing and analysis is by far the most common category used to describe the software. The prominence of data reduction and interaction, each considered applicable to 50\% of projects.
Interestingly, tools and infrastructure, modeling, and simulation were also similarly prominent (50\% and 47\%, respectively). ``Modeling and simulation" is a term often used to characterize applications in CSE, and may indicate fairly routine use of these techniques in the analysis of experimental and observational data. The number of projects categorized as tools and infrastructure was unexpectedly large. We plan to explore both of these categories in greater depth in follow-up studies.  It is interesting to note that looking out five years, nearly every category increases, suggesting an expectation that the projects will broaden in terms of the capabilities and thus more categories will apply in the future. The largest growth areas are expected to be in numerical libraries and data acquisition.

\begin{table}[tb]
\centering
\caption{Categories applicable to the focus software project.  The last two were provided as free-form additions to the list.}
\label{tab:11_27_categories}
\setlength{\tabcolsep}{1em}
\begin{tabular}{p{0.6\columnwidth}cc}
\hline\hline
 & Current & Expected \\
Category & Percentage & in Five Years \\
\hline
Data acquisition &  31\% & 47\% \\
Data reduction &  50\% & 55\% \\
Data processing and analysis &  74\% & 79\% \\
Data interaction (e.g., Jupyter notebooks, graphical or web interfaces, etc.) &  50\% & 55\% \\
Data dissemination or sharing & 21\% & 29\% \\
Modeling and simulation & 46\% & 60\% \\
Numerical libraries & 21\% & 45\% \\
Tools and infrastructure & 50\% & 55\% \\
Deep learning, machine learning, text analysis & 3\% & 3\% \\
Optimization & 3\% & 3\% \\
\hline\hline
\end{tabular}
\end{table}

We also asked several questions intended to elicit the importance of various technologies or approaches to the software projects.  The first question asked for an assessment of the importance of eight explicitly named technologies (on a 4-point scale), while the second questions was request for a free-form response.

Table~\ref{tab:12_28_technologies} shows the number and percentage of the 38 survey respondents who indicated that each technology was of moderate or higher importance (responses of 3 or 4).  This question was intended to gauge the extent of technologies  we thought might be ``emergent" in this community.  We were not surprised to see that continuous integration/deployment was essential to most of the projects (76\%), followed by numerical libraries (66\%) as many EOS efforts require fairly sophisticated numerical approaches. Data storage and interaction technologies were also important (55\% and 63\%, respectively). Cloud-based deployment is only relevant to roughly one-third of the respondents (32\%) and  only a small minority (5\%) rated cloud application programming interface (API) services as important, despite growing trends in cloud computing technologies in the last decades. This can be interpreted due to either lack of expertise in available cloud technologies, or that the cost of migrating operations might not justify the added value to the funded science deliverables. This is something we plan to explore further in follow-up studies. Looking ahead five years, all of the listed technologies are expected to increase in their importance to software projects. The largest growth is expected in the importance of cloud API services, with numerical libraries, data dissemination or sharing, and cloud deployment and operation technologies also significantly increased in importance.

\begin{table}[tb]
\centering
\caption{Technologies considered to be of moderate or higher importance to the software projects.}
\label{tab:12_28_technologies}
\setlength{\tabcolsep}{1em}
\begin{tabular}{p{0.6\columnwidth}cc}
\hline\hline
 & Current & Expected \\
Technology & Percentage & in Five Years \\
\hline
Data storage & 55\% & 66\% \\
Data interaction (e.g., Jupyter notebooks, graphical or web interfaces, etc.) & 63\% & 68\% \\
Data dissemination or sharing & 39\% & 58\% \\
Continuous integration/continuous deployment & 76\% & 79\% \\
Server-based deployment and operation & 45\% & 53\% \\
Cloud deployment and operation (using cloud resources to make the software available to users) & 32\% & 42\% \\
Numerical libraries (using library-based APIs as part of your software solution, for example BLAS, solvers, etc.) & 66\% & 82\% \\
Use of cloud API services (using cloud-based APIs as part of your software solution, for example, GCP Life Sciences API, Vision API, Trefle API, GBIF API, etc.) & 5\% & 37\% \\
\hline\hline
\end{tabular}
\end{table}

The free-form version of the question was included with the expectation that a wide range of technologies would be considered useful to different projects.  A total of 33 of the 38 respondents answered this question, listing a total of 69 distinct items, many appearing in multiple responses. For the sake of brevity, we have assigned the responses to categories, which are listed in Table~\ref{tab:13_technology_categories}.  We note that there is no unique and unambiguous way to categorize the tools and technologies named in the responses, but in this paper, our goal is to provide an overview; the specific responses are available in the survey dataset~\cite{raw_data}.  We note that application frameworks, libraries, and software development tools play significant roles in the software projects surveyed.  Python is also prominent, as are data tools, and artificial intelligence (AI) tools.

\begin{table}[tb]
\centering
\caption{Important technologies for the respondents' software projects, as categorized by the authors. There were a total of 69 unique technologies identified by the respondents.}
\label{tab:13_technology_categories}
\begin{tabular}{lcc}
\hline\hline
Technology Category & Responses & Percentage \\
\hline
AI tools & 6 & 9\% \\
Application Frameworks & 9 & 13\% \\
Build and test tools & 5 & 7\% \\
Computational Notebook Tools & 2 & 3\% \\
Data tools & 8 & 12\% \\
Deployment tools & 1 & 1\% \\
Hardware & 2 & 3\% \\
Libraries & 15 & 22\% \\
Python modules & 7 & 10\% \\
Software Development tools & 14 & 20\% \\
\hline\hline
\end{tabular}
\end{table}

We also asked respondents to indicate the programming languages used in their projects.  Overall, 37 of the 38 survey respondents answered this free-form question, naming a total of 58 distinct languages, 22 unique. Python and C\texttt{++} were the most prominent languages, cited by 78\% and 54\% of the responding projects, respectively. C and Javascript followed, listed in 16\% and 14\% of responses, respectively; Java and bash were the only other languages listed more than once (8\% and 5\%).  Programming for GPUs was noticeable as well, but diverse approaches were named (CUDA, HIP, Kokkos, OpenACC, OpenMP). 

Finally, in this section of the survey, we asked about the computational environments targeted by the project.  Respondents could select from five pre-defined responses as well as
being able to provide a free-form response (four were received).  The results are summarized in Table~\ref{tab:15_27_hardware}.  Here, we see that individual computers dominate, but in many cases, accelerators ({\it e.g.}, GPUs, or FPGAs) are used.  However usage of larger shared resources is also high.  About a third of project target cloud computing, which is consistent with the response to an earlier question in which a third of respondents also rated cloud deployment and operation technologies as important to their projects (Table~\ref{tab:12_28_technologies}). The use of GPU accelerators is consistent with the use of GPU programming languages as well.  Looking out five years, we see that respondents are generally expecting decreased use of single computers in favor of most other types of hardware, most significantly large-scale cloud and HPC environments.

\begin{table}[tb]
\centering
\caption{Hardware environments targeted by the projects.  Respondents could select any choices that applied, the last two entries were provided as free-form responses.}
\label{tab:15_27_hardware}
\setlength{\tabcolsep}{1em}
\begin{tabular}{p{0.55\columnwidth}cc}
\hline\hline
 & Current & Expected \\
Hardware Environment Target & Percentage & in Five Years\\
\hline
Single computer (laptop, desktop, or server) & 68\% & 60\% \\
Single computer with computational accelerator (e.g., GPU, FPGA, etc.) & 40\% & 50\% \\
Shared organizational resource (e.g., multiprocessor server or cluster, “edge” systems, etc.) & 63\% & 66\% \\
Lab-level, national, or commercial cloud resources (e.g., CADES cloud, AWS, etc.) & 34\% & 53\% \\
Lab-level, or national HPC resources (e.g., CADES condos, OLCF, etc.) & 47\% & 63\% \\
Neuromorphic processors & 3\% & 3\% \\
Quantum processors & 0\% & 3\% \\
Dedicated computational infrastructure for online analysis of experimental data (cpu + gpu +fpga) & 3\% & 0\% \\
Resources/infrastructure frozen for the duration of the experiment & 0\% & 3\% \\
\hline\hline
\end{tabular}
\end{table}

\subsection{Skills and Training}

This section gauges their assessment of their own knowledge and that of their project team as a whole in various, as well as the areas in which new knowledge would be important. We requested responses on a 4-point scale ranging from ``not knowledgeable" (1) to ``very knowledgeable" (4), or ``not important" (1) to ``very important" (4).  Table~\ref{tab:17_18_19_knowledge} summarizes the responses of 3-4 for personal knowledge, team knowledge, and the importance of acquiring new knowledge.  Respondents are fairly confident in both their own and their team's knowledge of most areas. The weakest area was computer hardware, which also scored the lowest in terms of the importance of improving knowledge. Respondents were somewhat more comfortable with their (and their teams') knowledge of the science, algorithms, and software tools for their work than with software development practices. However software development practices and algorithms rated slightly higher than the science and software tools in terms of areas needing more knowledge. 

\begin{table}[tb]
\centering
\caption{Self-assessed level of knowledge in various areas for the respondent and their team, and the perceived importance of gaining new knowledge in the area.}
\label{tab:17_18_19_knowledge}
\begin{tabular}{p{0.6\columnwidth}ccc}
\hline\hline
 & Personal & Team & New \\
Area & Knowledge & Knowledge & Knowledge \\
\hline
The scientific context of the software & 84\% & 95\% & 69\% \\
The algorithms and methods required to implement the software & 84\% & 84\% & 74\% \\
The software frameworks, libraries, and tools to support the implementation of the software & 79\% & 82\% & 71\% \\
Software development best practices & 68\% & 79\% & 74\% \\
The computer hardware to which the software is targeted (including cloud computing, if appropriate)  & 58\% & 66\% & 55\% \\
\hline\hline
\end{tabular}
\end{table}

\subsection{Preparing for the Future}

The final section of the survey asked respondents about their concerns about their projects in the next five years, and their overall confidence level in being able to deal with the changes they anticipate.

As we see in Table~\ref{tab:31_concerns}, the most significant area of concern has to do with the performance of the software, followed by changes in the hardware environment and the requirements for the maintainability or sustainability of the software.  Concerns about changes in the hardware environment here are noteworthy, given the fact that in Table~\ref{tab:17_18_19_knowledge} respondents were both least confident in their personal and their team's knowledge of the hardware, \emph{and} ranked it the least important area in which to gain more knowledge. The areas of least concern were the scientific context of the project (which was also an area of high confidence in Table~\ref{tab:17_18_19_knowledge}) and in cloud service APIs. The low concern about cloud service APIs may be explained by their low importance in Table~\ref{tab:12_28_technologies}. 

When asked how confident they were as individuals in their ability to deal with the changes anticipated over the next five years, a significant portion (63\%) responded that they were moderately confident (3 on a 4-point scale), and one-fifth of the respondents (19\%) were highly confident. When asked about their \emph{team's} ability to deal with the coming changes, the majority (79\%) are confident, with 42\% being moderately confident and 37\% highly confident.

\begin{table}[tb]
\centering
\caption{Levels of concern about various possible changes in their projects over the next five years.  Responses were on a 4-point scale from ``not concerned" (1) to ``very concerned" (4).  We summarize the responses of moderate or high concern (3-4).}
\label{tab:31_concerns}
\begin{tabular}{p{0.75\columnwidth}cc}
\hline\hline
Aspect & Responses & Percentage \\
\hline
Changes in the hardware environment & 23 & 61\% \\
Changes in the scientific context & 12 & 32\% \\
Changes in the required functionality of the software & 23 & 61\% \\
Changes in the required usability of the software & 21 & 55\% \\
Changes in the required maintainability or sustainability of the software & 23 & 61\% \\
Changes in the required performance of the software & 25 & 66\% \\
Changes in or increased importance of data storage technologies & 18 & 47\% \\
Changes in or increased importance of numerical libraries & 16 & 42\% \\
Changes in or increased importance of data interaction technologies (e.g., Jupyter notebooks, graphical or web interfaces, etc.) & 19 & 50\% \\
Changes in or increased importance of continuous  integration/continuous deployment technologies & 17 & 44\% \\
Changes in or increased importance of cloud for making the software available to users & 16 & 42\% \\
Changes in or increased importance of the use of cloud API service technologies & 13 & 34\% \\
\hline\hline
\end{tabular}
\end{table}